\documentclass{article}
\usepackage{spconf,amsmath,graphicx,hyperref}
\usepackage{siunitx}
\usepackage{comment}
\usepackage{bm} 
\usepackage{booktabs}
\usepackage{array}     
\usepackage{hyperref}

\usepackage{tikz}
\usetikzlibrary{shapes.geometric, arrows}

\tikzstyle{layer} = [rectangle, rounded corners, minimum width=3cm, minimum height=1cm, text centered, draw=black, fill=gray!20]
\tikzstyle{arrow} = [thick,->,>=stealth]

\usepackage{url}
\usepackage{hyperref}

\title{Deep Learning-Based Prediction of Energy Decay Curves from Room Geometry and Material Properties}
\name{Imran Muhammad, Gerald Schuller} 
\address{Technische Universität Ilmenau, Germany}
\begin{document}
%
\maketitle

\begin{abstract}

Accurate prediction of energy decay curves (EDCs) enables robust analysis of room acoustics and reliable estimation of key parameters. We present a deep learning framework that predicts EDCs directly from room geometry and surface absorption. A dataset of 6000 shoebox rooms with realistic dimensions, source–receiver placements, and frequency-dependent wall absorptions was synthesized. For each configuration we simulate room impulse responses (RIRs) using Pyroomacoustics and compute target EDCs. Normalized room features are provided to a long short-term memory (LSTM) network that maps configuration to EDC.
Performance is evaluated with mean absolute error (MAE) and root mean square error (RMSE) over time. We further derive early decay time (EDT), reverberation time (T20), and clarity index (C50) from predicted and target EDCs; close agreement is observed (e.g., EDT MAE 0.017 s, T20 MAE 0.021 s). The approach generalizes across diverse rooms and supports efficient room-acoustics modeling for early-stage design and real-time applications.

\end{abstract}
\section{Introduction}
\label{sec:intro}

Room acoustics plays a central role in shaping the auditory experience in enclosed environments such as concert halls, classrooms, offices, and virtual spaces. Key room acoustic parameters, such as reverberation times (T20), early decay time (EDT) and clarity (C50) provide quantitative insight into how sound behaves in a space and are critical for evaluating acoustic quality of the closed space. These parameters are often derived from room impulse responses (RIRs) or energy decay curves (EDCs), both of which describe how sound energy evolves over time after an acoustic excitation \cite{kuttruff2009room}. Traditionally, accurate estimation of such acoustic descriptors requires either detailed physical modeling (e.g., Ray Tracing, Image Source Method (ISM), and wave-based simulations) \cite{vorlander2020auralization} or time-consuming on-site measurements. Simplified empirical approaches, such as Sabine’s \cite{Millington1932} or Eyring’s \cite{Sette1933} formulas, are also used, though they rely on idealized assumptions and may be inaccurate in complex rooms. Fast geometrical acoustics software such as pysound \cite {schissler2011gsound}, ITA Geometrical Acoustics \cite{ITAGeometricalAcoustics}, RAVEN (Room Acoustics Simulation for Virtual Environments) \cite {schroder2011raven}, openRay \cite{openRay}, EVERTims \cite{Laine2009}, GSoundSIR \cite{GSoundSIR} and Pyroomacoustics \cite {8461310} provide an efficient alternative to physical measurements. However, these methods are based on approximations and struggle to capture wave-based phenomena accurately, particularly at low frequencies. Wave-based solvers offer greater accuracy but are often computationally intensive and unsuitable for real-time applications. This trade-off between speed and accuracy presents an opportunity for deep learning or data-driven approaches. Crucially, such models should be trained on high-quality data, ideally derived from physical measurements or wave-based simulations, to ensure reliability.
Recently, data-driven approaches utilizing deep learning have shown significant promise in modeling complex acoustic environments by learning representations directly from structured input data \cite{yu2019roomgeometryestimationroom}. Deep neural networks have been increasingly employed in room acoustics research, including the prediction of key parameters such as reverberation time and clarity indices \cite{meng2023predicting, 10.1121/10.0013575, 10.1121/10.0013416,9286412}, the estimation of room impulse responses from incomplete or noisy measurements \cite{kim2023generative}, and the modeling of energy decay functions through specialized neural network architectures. These advancements highlight the effectiveness of deep learning in capturing the intricate temporal and spatial characteristics of room acoustics, providing efficient and scalable alternatives to conventional physics-based modeling techniques.

In this study, we introduce a new approach that predicts EDCs from room features as an intermediate step, instead of directly predicting full room impulse responses. EDCs provide a more compact and stable representation of a room’s temporal energy decay and reverberation characteristics \cite{kuttruff2009room}, which makes them better suited for learning-based methods. In contrast, RIRs are high-dimensional signals with complex temporal and spectral details, making direct prediction more challenging and error-prone. Since EDCs also serve as the basis for computing important acoustic parameters \cite{vorlander2020auralization}, this strategy simplifies the modeling task while preserving the essential information required for an accurate room acoustics analysis. The synthesis of RIRs from predicted EDCs is a promising direction for our future studies, but it is not within the scope of this study.

We propose a deep learning-based framework that uses a Long Short-Term Memory (LSTM) neural network to predict EDCs directly from the geometric features of the room and the material absorption characteristics. LSTM networks are particularly well suited for time series prediction tasks due to their ability to capture long-term dependencies and temporal patterns, overcoming the limitations of standard recurrent neural networks (RNNs) such as the problem of vanishing gradients \cite{Hochreiter1997long, bengio1994learning}. Compared to convolutional neural networks (CNNs), which are more effective in spatial pattern recognition, LSTMs provide improved performance on sequential data, which is crucial for modeling the temporal decay nature of EDCs \cite{karim2018lstm}. Although alternative architectures such as Transformers offer competitive performance in sequence modeling, their higher computational complexity and data requirements make LSTMs a more efficient choice for this application \cite {vaswani2017attention, bai2018empirical}. This makes LSTM a robust and computationally feasible solution for EDCs predictions from room configuration parameters.

Our key contributions are as follows. We present a deep learning model capable of predicting EDCs directly from realistic room geometries and room material absorption characteristics, reducing the need for full physical-based simulations or impulse response measurements. Second, we introduce a framework in which the key acoustic parameters derived from the predicted EDCs are validated against those obtained from the simulated data, thereby demonstrating the precision and practical applicability of the proposed method. This proposal offers a foundation for rapid and efficient estimation of room acoustic behavior and has potential applications in early-stage acoustic design, real-time auralization.

\section{Methodology}
\label{sec:methodology}
We developed an LSTM-based deep learning framework to predict the EDCs of shoebox rooms using its geometric parameters and absorption properties as input features. The model is trained on a dataset generated through realistic combinations of room dimensions, source-receiver placements, and surface material absorptions. These features are normalized and structured before being fed into the LSTM network, which is designed to learn the mapping of room features to the temporal profile of the EDCs. The training process uses the mean squared error (MSE) as the loss function.

\subsection{Dataset and Room Simulation Parameters}
\label{ssec:dataset}
A data-set comprising $6000$ distinct room configurations was used as input room features, with each room defined by its length ($L$), width ($W$), height ($H$), source and receiver positions in three dimensions (i.e. $X$, $Y$, and $Z$), and frequency-dependent absorption coefficients averaged for all walls. These parameters were varied over a range of realistic values (see Table \ref{tab:simulated_room_ranges} ) to simulate various acoustic environments. For input data generation, we used the Pyroomacoustics library \cite{scheibler2018pyroomacoustics} to simulate a diverse set of shoebox-shaped rooms. Pyroomacoustics applies geometrical acoustics techniques, namely, the Image Source Method (ISM) and ray tracing, suitable for modeling high-frequency sound propagation. However, these methods do not account for wave-based phenomena. The dimensions of the rooms and the positions of the source and receiver were systematically varied and the surface absorption properties were sampled from a measured material database \cite{8461310}. These absorption coefficients are used to approximate the surface behavior, though they primarily account for energy loss and not phase-related effects. RIRs and the corresponding EDCs are synthesized using a hybrid simulation approach (ISM and ray tracing) within the Pyroomacoustics framework and saved as time series data at sampling rate of \SI{44.1}{\kilo\Hz}. While efficient, these methods are based on geometrical acoustics and are primarily valid at high frequencies, where wave-based effects are less significant.

\begin{figure}
    \centering
    \includegraphics[width=1\linewidth]{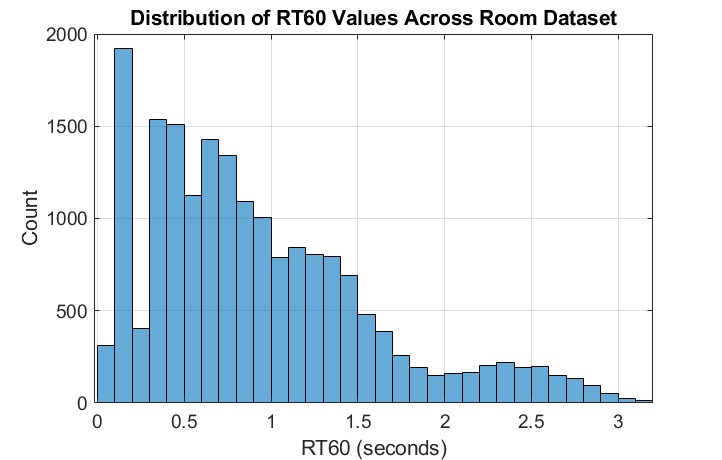}
    \caption{Histogram showing the distribution of $T60$ values across the room dataset}
    \label{fig:RTHist}
    \vspace{-0.5cm}
\end{figure}

To capture a wide range of acoustic environments, the reverberation time ($T60$) was computed from the RIR of each simulated room configuration. This ensures the dataset spans from nearly anechoic to highly reverberant conditions, covering both short and long $T60$ values. Figure \ref{fig:RTHist} shows the histogram of $T60$ across all configurations, highlighting the variability and coverage of reverberation conditions. Such diversity is crucial for training and evaluating models that must generalize across different room types.

\begin{table}[t]
\centering
\caption{Range of simulated room acoustic features used as input parameters for LSTM model training}
\label{tab:simulated_room_ranges}
\begin{tabular}{@{}>{\raggedright\arraybackslash}p{0.5\linewidth}c@{}}
\toprule
\textbf{Parameter} & \textbf{Range / Values} \\
\midrule
Dimensions ($L$ × $W$ × $H$)& $3-6$ m × $3-6$ m × $2.5-4$ m\\
Source-Receiver Distances& $1 - 4$ m\\
Wall Absorptions Range& $0.14 - 0.65$ ($125$ - $8000$ \SI{}{Hz})\\
Total Room Configurations& $6000$\\
\end{tabular}
\end{table}

\subsubsection{Input Data Preprocessing and Normalization}
\label{sssec:Data Preprocessing and Normalization}
The input features, representing room parameters, were scaled using \textit{MinMax()} normalization to the $[0, 1]$ range. Though, the normalized energy decay curves dataset inherently fall within this interval, \textit{MinMax} normalization was still applied as a standard preprocessing step to maintain consistency with best practices in deep learning and to promote stable training dynamics. Following normalization, the data were reshaped into a three-dimensional format of $[N, 1, 16]$, where $N$ denotes the number of rooms and $16$ corresponds to the number of features per room. While the input features are not temporally sequential, the Long Short-Term Memory (LSTM) architecture requires a three-dimensional input of the form (batch\_size, sequence\_length, feature\_dimension). To satisfy this requirement, each room sample was represented as a sequence of length one containing $16$ feature values. This transformation enables compatibility with the LSTM structure while leveraging its gating mechanisms to model potentially complex inter-dependencies among the input features. Although a two-dimensional input of shape $[N,16]$ would be sufficient for a fully connected neural network, the LSTM architecture mandates the inclusion of a sequence dimension, even if it represents only a single step. For model evaluation, the dataset was partitioned into $60\%$ training, $20\%$ validation, and $20\%$ test subsets.

\begin{figure}[h!]
    \centering
    \resizebox{0.35\textwidth}{0.35\textheight} {
    
    \begin{tikzpicture}[scale=1, node distance=1.5cm, every node/.style={font=\small}, thick]
    \node (input) [layer] {Input Sequence};
    \node (lstm) [layer, below of=input] {LSTM Layer (128 hidden units)};
    \node (drop1) [layer, below of=lstm] {Dropout (30\%)};
    \node (dense) [layer, below of=drop1] {Fully Connected Layer  (2048 neurons, ReLU)};
    \node (drop2) [layer, below of=dense] {Dropout (30\%)};
    \node (output) [layer, below of=drop2] {Output Layer  (target length, Linear)};
    \draw [arrow] (input) -- (lstm);
    \draw [arrow] (lstm) -- (drop1);
    \draw [arrow] (drop1) -- (dense);
    \draw [arrow] (dense) -- (drop2);
    \draw [arrow] (drop2) -- (output);

    \end{tikzpicture}
    }
    \caption{Architecture of the LSTM model used for EDC prediction.}
    \label{fig:lstm_model}
\end{figure}
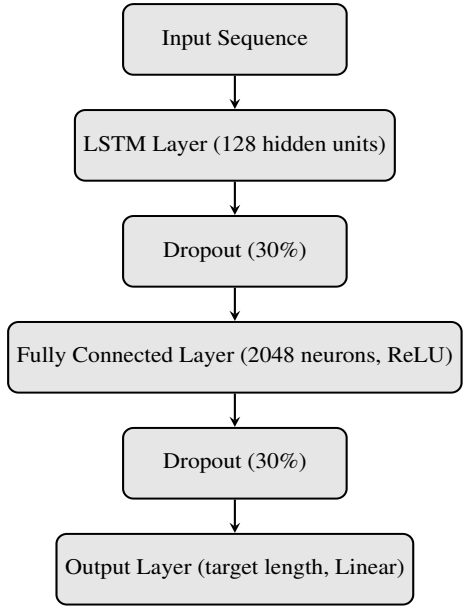

\subsection{LSTM Model Architecture}
\label{subsec:LSTM Model Architecture}

As Figure \ref{fig:lstm_model} shows, the LSTM model was implemented in PyTorch with a single LSTM layer of \SI{128}{} hidden units, followed by a \SI{30}{\%} dropout. A fully connected layer with \SI{2048}{} ReLU-activated neurons and another dropout layer precede the output layer, which has a size equal to the target EDC sequence and uses a linear activation for regression. The model was trained with the Adam optimizer ($LR = 0.001$) using mean squared error loss, with early stopping after \SI{10}{} epochs of no validation improvement. We compute MSE loss for EDCs on the linear scale instead of the dB scale. The dB transform exaggerates small variations in the late decay and downplays early decay differences, biasing the model. Linear loss better reflects true energy differences and yields a more stable, physically meaningful objective.

\section{Results and Discussion}
\label{sec:results}
After training, the model shows strong predictive accuracy on unseen rooms. Predicted EDCs are evaluated and used to derive room acoustic parameters (mentioned in Section \ref{sec:intro}). These are compared with those from target EDCs to assess reliability.

\begin{figure}
    \centering
    \includegraphics[width=1\linewidth]{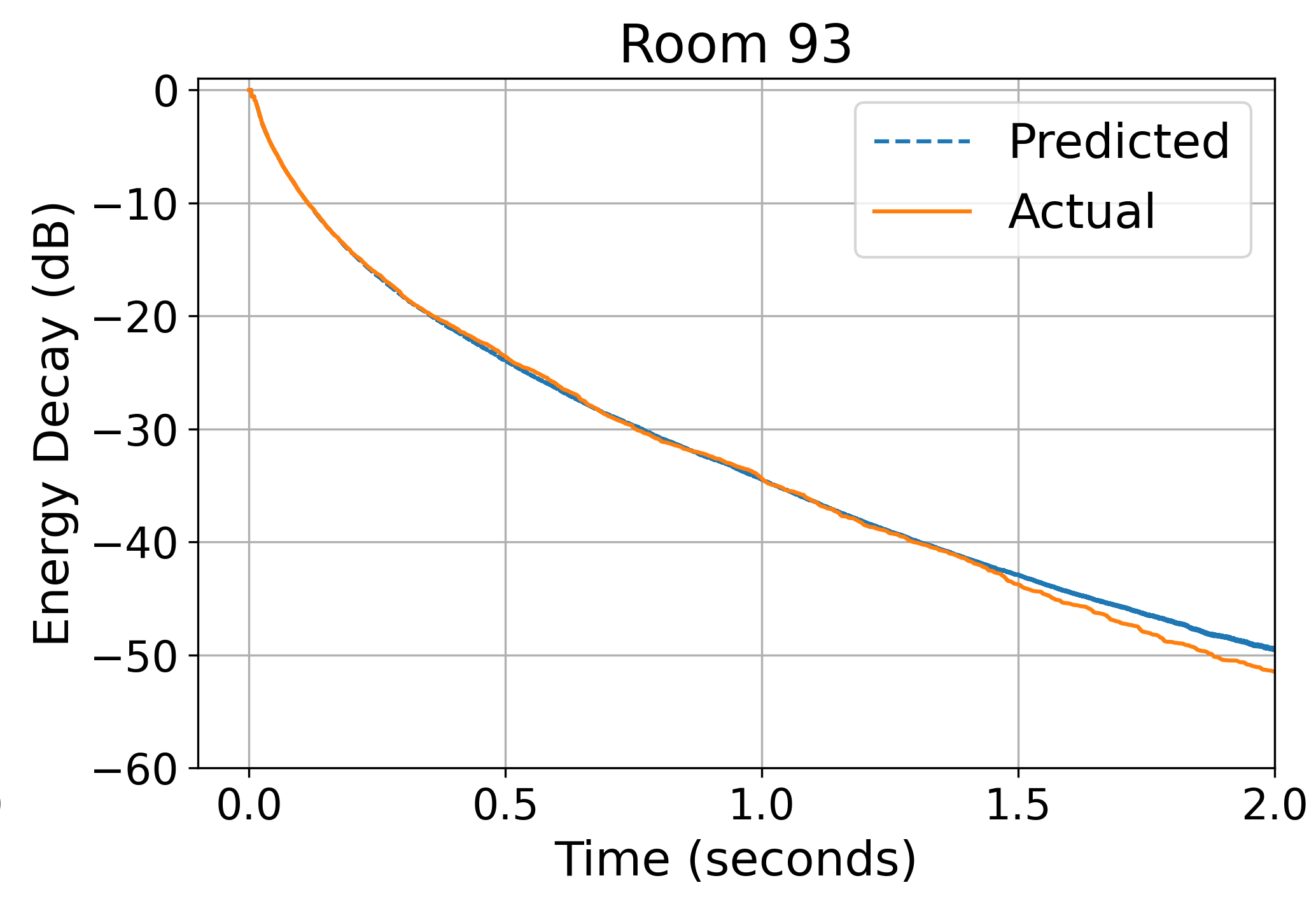}
    \caption{Target and predicted EDCs for randomly selected room}
    \label{fig:edcplot}
\end{figure}

\begin{figure}
    
    \centering
    \includegraphics[width=1\linewidth]{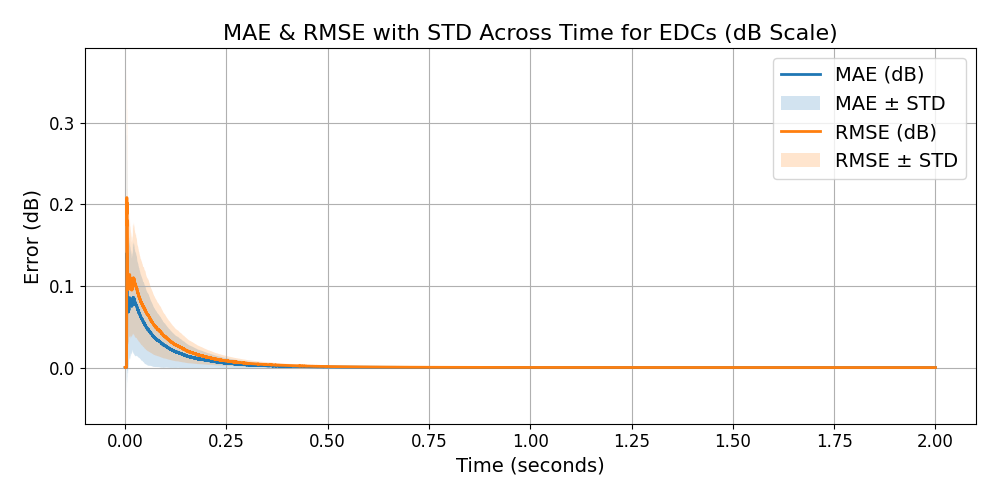}
    \caption{\textbf{MAE} and \textbf{RMSE} of the predicted EDCs, averaged over time across all room configurations}
    \label{fig:EDCError}
    \vspace{-0.5cm}
\end{figure}

\subsection{Prediction of Energy Decay Curves}
To assess the model's predictive performance, qualitative comparisons were conducted between the target (i.e. simulated) and predicted EDCs for a randomly selected room as shown in Figure ~\ref{fig:edcplot}. The graph reveal a strong agreement between the predicted and target EDC curves, suggesting that the model effectively captures the temporal decay characteristics of rooms. The predicted EDCs not only replicate the overall decay envelope but also capture fine-grained fluctuations in the decay profile. To further evaluate the accuracy of the predicted EDCs, we reported the MAE and RMSE computed across all predicted EDCs, averaged over time. Additionally, we included the standard deviation to reflect variability across the dataset as shown in Figure~\ref{fig:EDCError} to provide a more comprehensive evaluation of model behavior.

\begin{table}[t]
\centering
\caption{Performance of LSTM Model: \textbf{MAE}, \textbf{RMSE}, and coefficient of determination (\textbf{$R^2$}) computed between predicted and target EDCs}
\begin{tabular}{@{}lccc@{}}
\toprule
\textbf{Metric} & \textbf{MAE}& \textbf{RMSE}& \textbf{R\textsuperscript{2}} \\
\midrule
EDT (s)& 0.017& 0.023& 0.987\\
T20 (s)& 0.021& 0.029& 0.978\\
C50 (dB)& 0.917& 2.023& 0.888\\
\end{tabular}
\label{tab:acoustic_param_accuracy}
\end{table}

\subsection{Acoustic Parameter Estimation Accuracy}
\label{subsec:Acoustic Parameter Estimation Accuracy}
Additional analyses such as key room acoustic parameters (as discussed in Section~\ref{sec:intro}) were derived from both the target (simulated) and predicted EDCs. The predictive performance was quantified using MAE, RMSE, and $R^2$ (coefficient of determination) for each parameter across all room configurations in the test dataset. Table \ref{tab:acoustic_param_accuracy} summarizes the error metrics.

\begin{figure}
    \centering
    \vspace{-0.25cm}
    \includegraphics[width=1\linewidth]{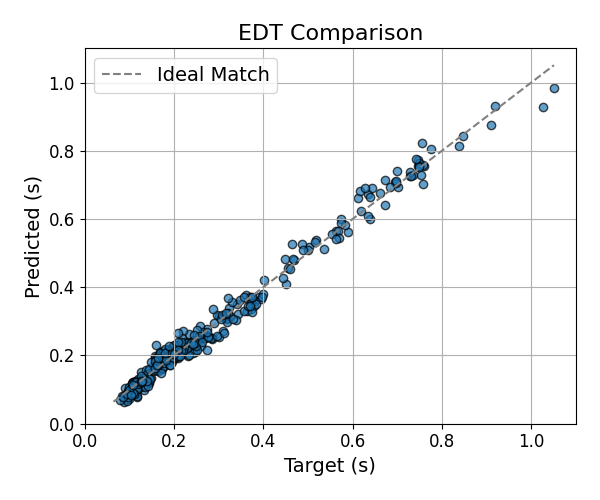}
    \caption{Comparison of predicted and target EDT values}
    \label{fig:EDT}
\end{figure}

\subsubsection{EDT and Reverberation Times (EDT, T20)}
\begin{figure}
    \centering
    \includegraphics[width=1\linewidth]{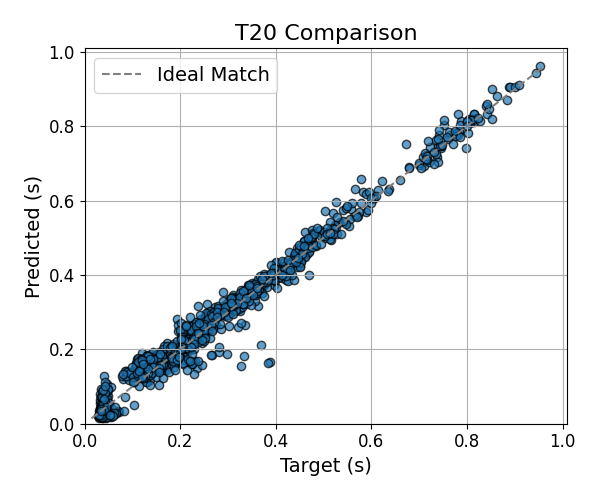}
    \caption{Comparison of predicted and target T20 values}
    \label{fig:T20}
\end{figure}

Figures \ref{fig:EDT} and \ref{fig:T20} compare the predicted vs. target values for each parameter. Across these graphs, a close alignment along the identity line ($y = x$) indicates strong agreement with the model. Particularly, the clustering around the ideal match line was the most prominent suggests that the model captured early reverberant characteristics effectively. The model shows a good performance in estimating EDT and T20, with a low MAE of \SI{0.017}{s} and an RMSE of \SI{0.023}{s} and MAE of \SI{0.021}{s} and an RMSE of \SI{0.029}{s} respectively. The high $R^2$ value of \SI{0.987}{} and \SI{0.978}{} for EDT and T20 respectively indicates that the predictions closely follow the trend of target values, confirming the model's ability to learn the early decay slope of the EDCs.

\section{Summary and outlook}
\label{sec:Summary and outlook}
The LSTM-based model demonstrated good performance in predicting EDCs and estimating key acoustic parameters with good accuracy. The model achieved low error values and consistently high $R^2$ scores. Our future work is on extending the model's generalization to real-world measured data and synthesis of full RIRs conditioned on these EDCs using methods such as stochastic modeling, inverse filtering, or generative neural approaches. We believe this approach enables accurate and plausible RIR synthesis for applications such as real-time room acoustics applications, such as intelligent room tuning, auralization engines, or speech enhancement systems, where fast and accurate acoustic parameter estimation is critical. The source code, preprocessed dataset, and trained model used in this study are available at \cite{AMSgithub}.

\vfill
\pagebreak

\bibliographystyle{IEEEbib}
\bibliography{strings,refs}

\end{document}